\documentclass[seceq]{ptptex}
\usepackage{wrapft}
\usepackage{graphicx}

\newcommand{\bra}[1]{\langle {#1} |}     
\newcommand{\ket}[1]{| {#1} \rangle}     
\newcommand{\rbra}[1]{( {#1} |}     
\newcommand{\rket}[1]{| {#1} )}     
\newcommand{\dbra}[1]{\langle {#1} |\!|}     
\newcommand{\dket}[1]{|\!| {#1} \rangle}     
\newcommand{\wtilde}[1]{\widetilde{#1}} 

\def\beq{\begin{eqnarray}}
\def\eeq{\end{eqnarray}}
\def\bsub{\begin{subequations}}
\def\esub{\end{subequations}}
\def\b{\begin{equation}}

\title{
On the Color-Singlet States in Many-Quark Model with the $su(4)$-Algebraic Structure. I
}
\subtitle{
Color-Symmetric Form
}    

\author{
Yasuhiko {\sc Tsue},$^{1}$ 
Constan\c{c}a {\sc Provid\^encia},$^{2}$ 
Jo\~ao da {\sc Provid\^encia}$^{2}$ and 
Masatoshi {\sc Yamamura}$^{3}$  
}

\inst{
$^{1}$Physics Division, Faculty of Science, Kochi University, Kochi 780-8520, 
Japan\\
$^{2}$Departamento de F\'{i}sica, Universidade de Coimbra, 3004-516 Coimbra, 
Portugal\\
$^{3}$Department of Pure and Applied Physics, 
Faculty of Engineering Science,\\
Kansai University, Suita 564-8680, Japan
\\

}

\recdate{
\today
}

\abst{
In the modified Bonn quark model in which quark-pairing and particle-hole 
interactions are included, color-singlet states are examined in addition to the 
color-neutral quark-triplet state by a method of boson realization. 
The region which consists of single-quarks, quark pairs and quark triplets 
is analyzed. 
}

\begin{document}

\maketitle

\section{Introduction}

It is believed from a theoretical viewpoint that an attractive quark-quark interaction 
leads to a quark pairing state, namely, a color superconducting state 
may exist in a quark matter at high density.\cite{01}
As an effective model including the quark pairing interaction, the Bonn quark model 
is an interesting one.\cite{02} 
This model has originally been introduced in order to describe 
a nucleus as a system of interacting quarks. 
This model actually has the interesting feature of the formation of the color-neutral 
quark-triplet as a nucleon and a $\Delta$-particle. 

Recently, the Bonn quark model has been reinevestigated by the present authors in Ref.\citen{1}, 
which is hereafter referred to as (A), in order to study possible states in the many-quark system. 
It was there indicated that the original Bonn quark model has an $su(4)$-dynamical 
symmetry. 
Thus, preserving only the color $su(3)$-symmetry, an $su(4)$-symmetry breaking term 
is introduced, which represents a particle-hole type interaction 
in terms of the quark shell model. 
We call this model the modified Bonn quark model. 
In (A), exact eigenstates were derived by using of the boson realization technique. 
Further, in Ref.\citen{2}, which is hereafter referred to as (B), exact eigenstates 
with single-quark, quark-pairing and quark-triplet states are 
treated in a unified way. 
Also, in Ref.\citen{3}, which is referred to as (C), the ground state was investigated 
in the modified Bonn quark model as a function of the quark number and 
the particle-hole interaction strength owing to the $su(4)$-symmetry breaking. 
It was shown that the quark-pairing state or the color-neutral quark-triplet state 
was realized in a certain parameter region including the transition region of both states.

In this paper, we examine a ``color-singlet" state in the 
modified Bonn quark model. 
In (A), (B) and (C), the color-neutral quark-triplet state was only realized 
as the color-singlet state. 
The purpose of this paper is to reinvestigate the color-singlet state. 
We will give a certain condition for a state to be a ``color-singlet" in the boson 
realization. 
It will be shown that this condition leads to the superposition of the 
eigenstates constructed on minimum weight states in the $su(3)$-subalgebras which are included 
in the $su(4)$-algebra governing the original Bonn quark model. 
And, we must point out that color-symmetric form of a modified Bonn quark model enables us to treat 
the present model in terms of the above-mentioned method. 
The form adopted in (A) has been not based on the color-symmetric form.

This paper is organized as follows: 
In the next section, an outline of the modified Bonn quark model is given 
and a concept of ``color-singlet" state is explained. 
In \S 3, the method of the boson realization in this model 
is given in a slightly different manner from that developed in (A), 
that is, the color-symmetric form. 
In \S 4, energy eigenstates are constructed on a single minimum weight state of 
$su(3)$-subalgebra. 
In \S 5, the energy eigenstates are extended and are constructed as a 
superposition based on the minimum weight states of possible $su(3)$-subalgebra. 
In \S 6, the ``color-singlet" state minimizing the eigenvalue of the 
$su(3)$-Casimir operator is examined and the regions which consist of 
single-quarks, quark-pairs and quark-triplets are depicted in a parameter space. 
The last section is devoted to a brief summary.

\section{Outline of the model}

The model treated in this paper has been already discussed in our 
three papers, (A), (B) and (C), 
in which, concerning the color-singlet states, we examined only the quark-triplet 
states. 
In this section, we will give outline of the model, including certain new aspects 
to be investigated. 
Of course, the new aspects will play a central role for the understanding of the 
color-singlet states in general cases. 
This model is formulated in terms of the $su(4)$-generators constructed by the 
bilinear forms of the quark creation and annihilation operators. 
The color quantum numbers are specified as $i=1$, 2 and 3. 
Each color state has the degeneracy $2\Omega$. 
Here, $2\Omega=2j_s+1$ and $j_s$ is a half-integer. 
Any single-particle state is specified as $(i,m)$ with $m=-j_s,\ -j_s+1,\cdots ,\ j_s-1$ 
and $j_s$, and the quark creation and annihilation operators in $(i,m)$ are 
denoted by $c_{im}^*$ and $c_{im}$, respectively. 
For simplicity, we neglect the degrees of freedom related to the isospin. 
We define the following bilinear forms for the quark creation and annihilation operators:
\bsub\label{2-1}
\beq
& &{\wtilde S}^1=\sum_m c_{2m}^*c_{3{\wtilde m}}^* \ , \qquad
{\wtilde S}^2=\sum_m c_{3m}^*c_{1{\wtilde m}}^*\ , \qquad
{\wtilde S}^3=\sum_m c_{1m}^*c_{2{\wtilde m}}^* \ , \nonumber\\
& &{\wtilde S}_1=({\wtilde S}^1)^* \ , \qquad\qquad\ 
{\wtilde S}_2=({\wtilde S}^2)^* \ , \qquad\qquad\ \ 
{\wtilde S}_3=({\wtilde S}^3)^* \ , 
\label{2-1a}\\
& &{\wtilde S}_1^2=\sum_m c_{2m}^*c_{1m} \ , \qquad
{\wtilde S}_2^3=\sum_m c_{3m}^*c_{2m} \ , \qquad
{\wtilde S}_3^1=\sum_m c_{1m}^*c_{3m} \ , \nonumber\\
& &{\wtilde S}_2^1=({\wtilde S}_1^2)^* \ , \qquad\qquad\ 
{\wtilde S}_3^2=({\wtilde S}_2^3)^* \ , \qquad\qquad\ \ 
{\wtilde S}_1^3=({\wtilde S}_3^1)^* \ , 
\label{2-1b}\\
& &{\wtilde S}_1^1=\sum_m (c_{2m}^*c_{2m}+c_{3m}^*c_{3m})-2\Omega\ , \qquad
{\wtilde S}_2^2=\sum_m (c_{3m}^*c_{3m}+c_{1m}^*c_{1m})-2\Omega\ , \nonumber\\
& &{\wtilde S}_3^3=\sum_m (c_{1m}^*c_{1m}+c_{2m}^*c_{2m})-2\Omega\ .
\label{2-1c}
\eeq
Here, $c_{i{\wtilde m}}^*=(-)^{j_s-m} c_{i -m}^*$. 
The form (\ref{2-1c}) gives us 
\beq\label{2-1d}
& &{\wtilde N}_i={\wtilde N}-{\wtilde S}_i^i -2\Omega \ , \qquad
{\wtilde N}=\sum_i {\wtilde N}_i \ .
\eeq
\esub
Here, ${\wtilde N}_i$ and ${\wtilde N}$ denote the quark number operator 
in the color $i$ and the total quark number operator, respectively. 
The operators defined in the relation (\ref{2-1}) are generators of the 
$su(4)$-algebra: 
\bsub\label{2-2}
\beq\label{2-2a}
& &[\ {\wtilde S}^i \ ,\ {\wtilde S}^j\ ]=0 \ , \qquad
[\ {\wtilde S}^i \ ,\ {\wtilde S}_j\ ]={\wtilde S}_i^j \ , \nonumber\\
& &[\ {\wtilde S}_i^j \ ,\ {\wtilde S}^k\ ]=\delta_{ij}{\wtilde S}^k + \delta_{jk}{\wtilde S}^i \ , 
\nonumber\\
& &[\ {\wtilde S}_i^j \ ,\ {\wtilde S}_l^k\ ]=\delta_{jl}{\wtilde S}_i^k - \delta_{ik}{\wtilde S}_l^j \ . 
\eeq
The Casimir operator of the $su(4)$-algebra, which we denote as ${\wtilde {\mib P}}^2$, 
is expressed in the form 
\beq\label{2-2b}
{\wtilde {\mib P}}^2&=&
\sum_i \left({\wtilde S}_i{\wtilde S}^i +{\wtilde S}^i{\wtilde S}_i\right)
+\sum_{i\neq j}{\wtilde S}_j^i{\wtilde S}_i^j \nonumber\\
&+&\frac{1}{4}
\left[\left({\wtilde S}_2^2-{\wtilde S}_3^3\right)^2+
\left({\wtilde S}_3^3-{\wtilde S}_1^1\right)^2+
\left({\wtilde S}_1^1-{\wtilde S}_2^2\right)^2\right] \ .
\eeq
\esub

As a sub-algebra, the $su(4)$-algebra contains the $su(3)$-algebra. 
The following six operators play a part of eight kinds of the 
$su(3)$-generators: 
\beq\label{2-3}
{\wtilde S}_1^2 \ , \quad 
{\wtilde S}_2^3 \ , \quad 
{\wtilde S}_3^1 \ , \quad 
{\wtilde S}_2^1 \ , \quad 
{\wtilde S}_3^2 \ , \quad 
{\wtilde S}_1^3 \ .
\eeq
Concerning the choice of the remaining two, formally, there exist six cases: 
\bsub\label{2-4}
\beq
& &\frac{1}{2}({\wtilde S}_2^2-{\wtilde S}_3^3) \ , \qquad
{\wtilde S}_1^1-\frac{1}{2}({\wtilde S}_2^2+{\wtilde S}_3^3) \ ,
\label{2-4a}\\
& &\frac{1}{2}({\wtilde S}_3^3-{\wtilde S}_1^1) \ , \qquad
{\wtilde S}_2^2-\frac{1}{2}({\wtilde S}_3^3+{\wtilde S}_1^1) \ ,
\label{2-4b}\\
& &\frac{1}{2}({\wtilde S}_1^1-{\wtilde S}_2^2) \ , \qquad
{\wtilde S}_3^3-\frac{1}{2}({\wtilde S}_1^1+{\wtilde S}_2^2) \ ,
\label{2-4c}
\eeq
\esub
\vspace{-0.5cm}
\bsub\label{2-5}
\beq
& &\frac{1}{2}({\wtilde S}_3^3-{\wtilde S}_2^2) \ , \qquad
{\wtilde S}_1^1-\frac{1}{2}({\wtilde S}_3^3+{\wtilde S}_2^2) \ ,
\label{2-5a}\\
& &\frac{1}{2}({\wtilde S}_1^1-{\wtilde S}_3^3) \ , \qquad
{\wtilde S}_2^2-\frac{1}{2}({\wtilde S}_1^1+{\wtilde S}_3^3) \ ,
\label{2-5b}\\
& &\frac{1}{2}({\wtilde S}_2^2-{\wtilde S}_1^1) \ , \qquad
{\wtilde S}_3^3-\frac{1}{2}({\wtilde S}_2^2+{\wtilde S}_1^1) \ .
\label{2-5c}
\eeq
\esub
By transposing ${\wtilde S}_2^2$ and ${\wtilde S}_3^3$, the case (\ref{2-5a}) is 
obtained from the case (\ref{2-4a}). 
The cases (\ref{2-4b}) and (\ref{2-4c}) and also the cases (\ref{2-5b}) and 
(\ref{2-5c}) are obtained from the cases (\ref{2-4a}) and (\ref{2-5a}), 
respectively, by the cyclic permutation $(1\rightarrow 2\ ,\ 
2\rightarrow 3,\ 3\rightarrow 1)$. 
Usually, the eight $su(3)$-generators are composed of three $su(2)$-generators and 
its scalar and two sets of spinors. 
For each case, we can express the $su(3)$-generators in terms of 
$({\wtilde S}_i^j)$. 
In Appendix A, an example is shown for the case (\ref{2-4a}). 
For these six cases, the Casimir operator, which we denote as 
${\wtilde {\mib Q}}^2$, can be expressed in the form 
\beq\label{2-6}
{\wtilde {\mib Q}}^2&=&
\sum_{i\neq j}{\wtilde S}_j^i{\wtilde S}_i^j+\frac{1}{3}
\left[\left({\wtilde S}_2^2-{\wtilde S}_3^3\right)^2
+\left({\wtilde S}_3^3-{\wtilde S}_1^1\right)^2
+\left({\wtilde S}_1^1-{\wtilde S}_2^2\right)^2\right] \nonumber\\
&=&
\sum_{i\neq j}{\wtilde S}_j^i{\wtilde S}_i^j 
+\sum_{i}({\wtilde S}_i^i)^2-\frac{1}{3}\left(\sum_i{\wtilde S}_i^i\right)^2 \ .
\eeq
Naturally, the form (\ref{2-6}) is common to the six cases. 
It should be noted that the treatment in (A) $\sim$ (C) has been 
restricted only to the case (\ref{2-4a}). 
In this paper, we examine the remaining five cases on an equal footing with the case 
(\ref{2-4a}).

The Hamiltonian of the present model is expressed in the following form:
\beq\label{2-7}
{\wtilde H}_m={\wtilde H}+\chi {\wtilde {\mib Q}}^2 \ , \qquad
{\wtilde H}=-\sum_{i}{\wtilde S}^i{\wtilde S}_i \ .
\eeq
Here, $\chi$ denotes a real parameter. 
In the original model, which is usually called as the Bonn model, only 
the term ${\wtilde H}$ is taken into account. 
In (A) $\sim$ (C), we added new term $\chi{\wtilde {\mib Q}}^2$, which, as will be shown in 
\S 6, plays a fundamental role. 
The Hamiltonian ${\wtilde H}_m$ satisfies the relation 
\bsub\label{2-8}
\beq
& &[\ {\wtilde H}_m \ , {\wtilde S}_j^i\ ]=0\quad {\rm for}\quad i,\ j=1,\ 2\ {\rm and}\ 3 \ , 
\label{2-8a}\\
& &{\wtilde H}_m=\frac{1}{2}(1+2\chi){\wtilde {\mib Q}}^2-\frac{1}{2}\left({\wtilde {\mib P}}^2
-{\wtilde \Sigma}\right) \ , \nonumber\\
& &{\wtilde \Sigma}=\frac{1}{12}\left(\sum_i{\wtilde S}_i^i\right)\left(\sum_i{\wtilde S}_i^i-12\right)
=\frac{1}{3}(3\Omega-{\wtilde N})(3\Omega-{\wtilde N}+6) \ . 
\label{2-8b}
\eeq
\esub
The relation (\ref{2-8}) indicates that the present model has the $su(3)$ color symmetry. 
It may be interesting to solve the eigenvalue equation for ${\wtilde H}_m$ and 
analyze the present model in terms of focusing on the $su(3)$ color-symmetry.

In relation to quantum chromodynamics (QCD), the eigenstate of ${\wtilde H}_m$ should obey the condition for 
the color singlet:
\bsub\label{2-9}
\beq
& &{\wtilde S}_i^j \rket{cs}=0\quad {\rm for}\quad i\neq j \ , 
\label{2-9a}\\
& &{\wtilde S}_1^1\rket{cs}={\wtilde S}_2^2\rket{cs}={\wtilde S}_3^3\rket{cs} \ .
\label{2-9b}
\eeq
\esub
Here, $\rket{cs}$ denotes an eigenstate of ${\wtilde H}_m$. 
The condition (\ref{2-9}) gives us 
\bsub\label{2-10}
\beq\label{2-10a}
{\wtilde {\mib Q}}^2\rket{cs}=0 \ .
\eeq
Further, the relations (\ref{2-1c}), (\ref{2-1d}) and (\ref{2-9b}) lead us to 
\beq\label{2-10b}
{\wtilde N}_1\rket{cs}={\wtilde N}_2\rket{cs}={\wtilde N}_3\rket{cs}=\frac{N}{3}\rket{cs} \ .
\eeq
\esub
Here, $N$ denotes the total quark number in the state $\rket{cs}$. 
From the relation (\ref{2-10b}), we can learn that, only in the case 
where $N$ is a multiple of 3, the color-singlet 
state appears. 
All the states except $\rket{cs}$ are not color-singlet. 
Then, it may be desirable for the present model 
to expand the concept of the color-singlet state by requiring that the condition 
(\ref{2-9}) is satisfied by the state $\rket{cs}$ in the average.

Preparatory to accomplishment of the above aim, first, we consider the state 
$\rket{cs}_L$ which obeys the condition 
\bsub\label{2-11}
\beq
& &\rbra{cs}{\wtilde S}_i^j\rket{cs}_L=0 \quad {\rm for} \quad i\neq j \ , 
\label{2-11a}\\
& &\rbra{cs}{\wtilde S}_1^1\rket{cs}_L=\rbra{cs}{\wtilde S}_2^2\rket{cs}_L=\rbra{cs}{\wtilde S}_3^3\rket{cs}_L \ .
\label{2-11b}
\eeq
\esub
As can be seen in the relation (\ref{2-6}), ${\wtilde {\mib Q}}^2$ is of the 
positive-definite quadratic form for the $su(3)$-generators and, then, we have 
\bsub\label{2-12}
\beq
\rbra{cs}{\wtilde {\mib Q}}^2\rket{cs}_L \geq 0 \ .
\label{2-12a}
\eeq
Further, we have 
\beq\label{2-11c}
\rbra{cs}{\wtilde N}_1\rket{cs}_L=\rbra{cs}{\wtilde N}_2\rket{cs}_L=\rbra{cs}{\wtilde N}_3\rket{cs}_L=\frac{N}{3} \ .
\eeq
\esub
Since $\rket{cs}_L$ is generally not the eigenstate of ${\wtilde N}_1$, ${\wtilde N}_2$ and 
${\wtilde N}_3$, $N$ is not restricted to be a multiple of 3. 
The case $\rbra{cs}{\wtilde {\mib Q}}^2\rket{cs}=0$ gives us 
${\wtilde {\mib Q}}^2\rket{cs}=0$, which means 
$\rket{cs}_L=\rket{cs}$. 
It may be self-evident that $\rket{cs}$ is a member of the set 
$\{\rket{cs}_L\}$ and rather many of $\rket{cs}_L$ are not 
color-singlet states $\rket{cs}$. 
But, they leave the trace of the color-singlet states in the sense of the 
relation (\ref{2-11}).

The condition (\ref{2-9}) seems to be rather restrictive. 
Actually, it may be seldom to find effective theories of QCD, in which any state obeys the 
condition (\ref{2-9}) exactly. 
The present model is also not the exception. 
Then, regarding the condition (\ref{2-11}) as an approximation to the condition (\ref{2-9}), 
in this paper, we will investigate the state $\rket{cs}_L$, which, hereafter, 
we denote as the ``color-singlet" state.

Further, in this case, we supplement the following requirement: 
It may be desirable to search the state $\rket{cs}_L$, in which the expectation 
value $\rbra{cs}{\wtilde {\mib Q}}^2\rket{cs}_L$ is as small as possible, 
because, in some sense, $\rbra{cs}{\wtilde {\mib Q}}^2\rket{cs}_L$ plays a role of the 
standard deviation for the expectation values (\ref{2-11}). 
We denote the state $\rket{cs}_L$ obeying this requirement as $\rket{cs}_M$. 
From the next section, we will search an idea for constructing the 
counterparts of $\rket{cs}_L$ and $\rket{cs}_M$ presented in the 
Schwinger boson representation. 
We denote them as $\ket{cs}_L$ and $\ket{cs}_M$, respectively.

\section{Boson realization}

Instead of solving our problem directly in the original fermion space, in (A) $\sim$ (C), 
we adopt the idea of the boson realization. 
As was shown in Ref.\citen{b}, there exist infinite possibilities 
for the boson realization of the $su(4)$-algebra. 
As a possible choice, we formulated a possible form of the boson realization in terms of 
the following form:
\beq\label{3-1}
& &{\wtilde S}^i \rightarrow {\hat S}^i={\hat a}_i^*{\hat b}-{\hat a}^*{\hat b}_i \ , \qquad
{\wtilde S}_i \rightarrow {\hat S}_i={\hat b}^*{\hat a}_i-{\hat b}_i^*{\hat a} \ , \nonumber\\
& &{\wtilde S}_i^j \rightarrow {\hat S}_i^j=({\hat a}_i^*{\hat a}_j-{\hat b}_j^*{\hat b}_i)
+\delta_{ij}({\hat a}^*{\hat a}-{\hat b}^*{\hat b}) \ . 
\eeq
Here, $({\hat a}_i,{\hat a}_i^*)$, $({\hat b}_i,{\hat b}_i^*)$, $({\hat a},{\hat a}^*)$ 
and $({\hat b},{\hat b}^*)$ ($i=1,2,3)$ denote the eight kinds of boson operators. 
In (A), we have discussed the reason why the representation (\ref{3-1}) is acceptable. 
In relation to the representation (\ref{3-1}), we can define the $su(1,1)$-algebra 
in the present boson space: 
\beq\label{3-2}
& &{\hat T}_+=\sum_i{\hat b}_i^*{\hat a}_i^*+{\hat b}^*{\hat a}^* \ , \qquad
{\hat T}_-=\sum_i{\hat a}_i{\hat b}_i+{\hat a}{\hat b} \ , \nonumber\\
& &{\hat T}_0=\frac{1}{2}\sum_i({\hat b}_i^*{\hat b}_i+{\hat a}_i^*{\hat a}_i)
+\frac{1}{2}({\hat b}^*{\hat b}+{\hat a}^*{\hat a})+2 \ . 
\eeq
They satisfy 
\bsub\label{3-3}
\beq
& &[\ {\hat T}_+\ , \ {\hat T}_-\ ]=-2{\hat T}_0 \ , \qquad
[\ {\hat T}_0\ , \ {\hat T}_{\pm}\ ]=\pm {\hat T}_{\pm} \ ,
\label{3-3a}\\
& &[\ {\rm any\ of\ }({\hat T}_{\pm,0})\ , \ {\rm any\ of\ } ({\hat S}^i, \ {\hat S}_i,\ {\hat S}_i^j)\ ] 
=0 \ . 
\label{3-3b}
\eeq
\esub
We cannot find any counterpart of $({\hat T}_{\pm,0})$ in the original fermion space.
Since, in (A) $\sim$ (C), we discussed the role of the $su(1,1)$-algebra 
in our model, in this paper, we will not repeat its discussion in detail.

With the use of the correspondence (\ref{3-1}), the Hamiltonian in the 
boson space which comes from ${\wtilde H}_m$ given in the relation (\ref{2-7}) is 
obtained: 
\beq\label{3-4}
{\wtilde H}_m \rightarrow {\hat H}_m \ , \qquad
{\wtilde H} \rightarrow {\hat H}\ , \qquad 
{\wtilde {\mib Q}}^2 \rightarrow {\hat {\mib Q}}^2 \ .
\eeq
By solving the eigenvalue equation for ${\hat H}_m$, we are able to get various 
informations on the boson system. 
However, through these informations, we do not obtain any knowledge on the 
original fermion system. 
The reason is very simple: 
The present boson system does not contain the quantity related to $\Omega$, 
which characterizes the many-fermion model under investigation. 
In (A), we showed that the $c$-number $\Omega$ in the original fermion space is defined in terms of 
the $q$-number ${\hat \Omega}$ in the boson space: 
\beq\label{3-5}
\Omega \rightarrow {\hat \Omega}=n_0+\frac{1}{2}({\hat a}^*{\hat a}+{\hat b}^*{\hat b})
+\frac{1}{2}\sum_i ({\hat a}_i^*{\hat a}_i+{\hat b}_i^*{\hat b}_i) \ .
\eeq
Here, $n_0$ is a certain $c$-number which plays a role similar to the 
seniority number in the $su(2)$-pairing model. 
In the original paper of the Bonn model, $n_0$ is treated as the number of the 
$\Delta$-particles. 
In this paper, we make the positions of ${\hat \Omega}$ and ${n}_0$ reverse, that is, 
we regard ${\hat \Omega}$ and $n_0$ as $c$- and $q$-numbers, respectively. 
Then, ${\hat n}_0$ is expressed as 
\beq\label{3-6}
{\hat n}_0=\Omega-\frac{1}{2}({\hat a}^*{\hat a}+{\hat b}^*{\hat b})
-\frac{1}{2}\sum_i ({\hat a}_i^*{\hat a}_i+{\hat b}_i^*{\hat b}_i) \ .
\eeq 
In relation to ${\hat n}_0$, we introduce the operator ${\hat n}$ defined as 
\beq\label{3-7}
{\hat n}=\Omega+\frac{1}{2}({\hat a}^*{\hat a}-{\hat b}^*{\hat b})
-\frac{1}{2}\sum_i ({\hat a}_i^*{\hat a}_i-{\hat b}_i^*{\hat b}_i) \ .
\eeq 
Later, we will mention the meaning of ${\hat n}$. 
Further, in (A), we introduced the operator ${\hat N}$ which comes from the total 
quark number. 
Through the above mentioned reversal, ${\hat N}$ can be expressed in the form
\beq\label{3-8}
{\hat N}=3\Omega+\frac{3}{2}({\hat a}^*{\hat a}-{\hat b}^*{\hat b})
+\frac{1}{2}\sum_i({\hat a}_i^*{\hat a}_i-{\hat b}_i^*{\hat b}_i) \ . 
\eeq
Associated with ${\hat N}$, the quark number operator ${\hat N}_i$ in the color $i$ is 
expressed as 
\beq\label{3-9}
{\hat N}_i=\Omega+\frac{1}{2}({\hat a}^*{\hat a}-{\hat b}^*{\hat b})
+\frac{1}{2}\sum_j({\hat a}_j^*{\hat a}_j-{\hat b}_j^*{\hat b}_j)
-({\hat a}_i^*{\hat a}_i-{\hat b}_i^*{\hat b}_i) \ . 
\eeq
Of course, we have ${\hat N}=\sum_i{\hat N}_i$. 
The Casimir operator ${\hat {\mib Q}}^2$ can be expressed as 
\beq\label{3-10}
{\hat {\mib Q}}^2=\sum_{ij}({\hat a}_j^*{\hat a}_i-{\hat b}_i^*{\hat b}_j)({\hat a}_i^*{\hat a}_j
-{\hat b}_j^*{\hat b}_i)
-\frac{1}{3}\left(\sum_i({\hat a}_i^*{\hat a}_i-{\hat b}_i^*{\hat b}_i)\right)^2 \ .
\eeq
We must stress that the operators ${\hat n}_0$, ${\hat n}$, 
${\hat N}$ and ${\hat {\mib Q}}^2$ are color-symmetric and commute with ${\hat S}_i^j$.

As a possible idea for expressing the eigenstates of ${\hat H}_m$, in (B), 
we introduced the following operators: 
\beq
& &{\hat q}^i={\hat b}_i^*{\hat b}-{\hat a}^*{\hat a}_i \ , \qquad
{\hat q}_i={\hat b}^*{\hat b}_i-{\hat a}_i^*{\hat a} \ , 
\label{3-11}\\
& &{\hat B}^*=\sum_{i}{\hat q}^i{\hat S}^i \ , \qquad
{\hat B}=\sum_i{\hat S}_i{\hat q}_i \ .
\label{3-12}
\eeq
The commutation relations between $({\hat n}_0,{\hat n},{\hat N})$ and 
$({\hat q}^i,{\hat S}^i,{\hat B}^*)$ are given in the form
\beq
& &[\ {\hat n}_0 \ , \ {\hat q}^i\ ]=0 \ , \quad\ \ 
[\ {\hat n}_0 \ , \ {\hat S}^i\ ]=0 \ , \qquad
[\ {\hat n}_0 \ , \ {\hat B}^*\ ]=0 \ , 
\label{3-13}\\
& &[\ {\hat n} \ , \ {\hat q}^i\ ]={\hat q}^i \ , \quad\ \ \!\  
[\ {\hat n} \ , \ {\hat S}^i\ ]=0 \ , \qquad\ 
[\ {\hat n} \ , \ {\hat B}^*\ ]={\hat B}^* \ , 
\label{3-14}\\
& &[\ {\hat N} \ , \ {\hat q}^i\ ]={\hat q}^i \ , \quad\ \  
[\ {\hat N} \ , \ {\hat S}^i\ ]=2{\hat S}^i \ , \quad
[\ {\hat N} \ , \ {\hat B}^*\ ]=3{\hat B}^* \ . 
\label{3-15}
\eeq
The relation (\ref{3-15}) may be interesting. 
The operators ${\hat q}^i$, ${\hat S}^i$ and ${\hat B}^*$ carry one, two and three 
quarks, respectively. 
In (B), we mentioned that ${\hat q}^i$, ${\hat S}^i$ and ${\hat B}^*$ represent 
single-quark, quark-pair and quark-triplet, respectively. 
Concerning the relations to the $su(3)$-generators, we have the 
following commutation relations: 
\beq
& &[\ {\hat S}_i^j \ , \ {\hat S}^k \ ]=\delta_{jk}{\hat S}^i+\delta_{ij}{\hat S}^k \ , 
\label{3-16}\\
& &[\ {\hat S}_i^j \ , \ {\hat q}^k \ ]=-\delta_{ik}{\hat q}^j+\delta_{ij}{\hat q}^k \ , 
\label{3-17}\\
& &[\ {\hat S}_i^j \ , \ {\hat B}^* \ ]=2\delta_{ij}{\hat B}^* \ . 
\label{3-18}
\eeq
Through the relation (\ref{3-18}), we can see that ${\hat B}^*$ is a color-singlet 
operator.

The above is a reformulation of the boson realization for the  present 
many-quark model which has been discussed in (A) $\sim$ (C). 
We do not know the operator which plays the same role as that 
of ${\hat q}^i$ in the original fermion space. 
Even if it is possible, the form may be too complicated to handle it easily. 
This is a basic merit for treating the present fermion model in the 
boson space.

\section{Energy eigenstates constructed on a chosen single minimum weight state}

First, we will show the orthogonal set discussed in (A)$\sim$ (C). 
The discussion in (A) is based on the case (\ref{2-4a}) in which the 
$su(3)$-generators are defined in the form (\ref{a3}). 
Then, for the minimum weight state $\ket{m_1}$, we set up the following relations: 
\bsub\label{4-1}
\beq
& &{\hat S}_1\ket{m_1}={\hat S}_2\ket{m_1}={\hat S}_3\ket{m_1}=0 \ , 
\label{4-1a}\\
& &{\hat S}_2^1\ket{m_1}={\hat S}_3^1\ket{m_1}={\hat S}_3^2\ket{m_1}=0 \ , 
\label{4-1b}\\
& &{\hat S}_1^1\ket{m_1}=-2\sigma\ket{m_1}\ , \qquad
{\hat S}_2^2\ket{m_1}=-2\sigma_0\ket{m_1}\ , \qquad
{\hat S}_3^3\ket{m_1}=-2\sigma_0'\ket{m_1}\ . \quad
\label{4-1c}
\eeq
The relation (\ref{4-1c}) leads us to 
\beq
& &\left({\hat S}_1^1-\frac{1}{2}({\hat S}_2^2+{\hat S}_3^3)\right)\ket{m_1}
=-2\left(\sigma-\frac{1}{2}(\sigma_0+\sigma_0')\right)\ket{m_1} \ , \nonumber\\
& &\frac{1}{2}({\hat S}_2^2-{\hat S}_3^3)\ket{m_1}=-(\sigma_0-\sigma_0')\ket{m_1} \ .
\label{4-1d}
\eeq 
\esub
As a relation which has no counterpart in the original fermion space, we require 
\beq\label{4-2}
{\hat T}_-\ket{m_1}=0 \ .
\eeq
Under the expressions (\ref{3-1}) and (\ref{3-2}), the relations (\ref{4-1}) 
and (\ref{4-2}) gives the following explicit form for $\ket{m_1}$: 
\beq\label{4-3}
\ket{m_1}=({\hat b}_1^*)^{2(\sigma-\sigma_0)}({\hat b}^*)^{2\sigma_0}\ket{0} \ . \quad
(\sigma_0'=\sigma_0)
\eeq
It is noted that the boson realization (\ref{3-1}) leads to the solution 
$\sigma_0'=\sigma_0$ and in (A), we used the notation $\sigma_1$ for $\sigma$. 
The state $\ket{m_1}$ is specified by two quantum numbers $\sigma_0$ and 
$\sigma$. 
The relations (\ref{4-1b}) and (\ref{4-1d}) tell that $\ket{m_1}$ 
is also a minimum weight state of the $su(3)$-algebra and the eigenvalues shown 
in the relation (\ref{4-1d}) are 
$-2(\sigma-(\sigma_0+\sigma_0')/2)=-2(\sigma-\sigma_0)$ and 
$-(\sigma_0-\sigma_0')=0$, respectively. 
As is shown in the relation 
(\ref{a2a}) and (\ref{a3a}),  
${\hat I}_+={\hat S}_2^3$, ${\hat I}_-={\hat S}_3^2$ and 
${\hat I}_0=({\hat S}_2^2-{\hat S}_3^3)/2$ form the $su(2)$-algebra and 
$\ket{m_1}$ is a state with the $su(2)$-spin$=0$. 
By operating ${\hat S}^1$, ${\hat S}^2$, ${\hat S}^3$, 
${\hat S}_1^2$, ${\hat S}_1^3$ and ${\hat S}_2^3$, together with 
${\hat S}_1^1$, ${\hat S}_2^2$ and ${\hat S}_3^3$ on $\ket{m_1}$ appropriately, 
we obtain the eigenstates of ${\hat H}_m$. 
For example, we have 
\bsub\label{4-4}
\beq
& &\ket{\lambda\rho\sigma_0\sigma}=({\hat S}^3)^{2\lambda}({\hat S}^4)^{2\rho}\ket{m_1} \ , 
\label{4-4a}\\
& &{\hat S}^4={\hat S}^1\left({\hat S}_1^1-\frac{1}{2}({\hat S}_2^2+{\hat S}_3^3)\right)
+{\hat S}^2{\hat S}_1^2+{\hat S}^3{\hat S}_1^3 \ . 
\label{4-4a}
\eeq
\esub
In (A) and (B), we showed the expression (\ref{4-4}) in the relations 
(A$\cdot$4$\cdot$24) and (B$\cdot$5$\cdot$1) in notations slightly different from the above.

Next, we investigate the physical meaning and the role of the operators 
${\hat n}_0$ and ${\hat n}$. 
It is easily verified that $\ket{m_1}$ is an eigenstate of ${\hat n}_0$ and ${\hat n}$: 
\bsub\label{4-5}
\beq
& &{\hat n}_0\ket{m_1}=n_0\ket{m_1} \ , \quad n_0=\Omega-\sigma\ , \quad{\rm i.e.,}\quad 
\sigma=\Omega-n_0 \ , 
\label{4-5a}\\
& &{\hat n}\ket{m_1}=n\ket{m_1} \ , \quad n=\Omega+\sigma-2\sigma_0\ , \quad{\rm i.e.,}\quad 
\sigma_0=\Omega-\frac{1}{2}(n_0+n) \ . 
\label{4-5b}
\eeq
\esub
On the other hand, we have 
\beq\label{4-6}
{\hat N}_1\ket{m_1}=n\ket{m_1} \ , \qquad
{\hat N}_2\ket{m_1}={\hat N}_3\ket{m_1}=n_0\ket{m_1}\ .  
\eeq
We can learn in the relation (\ref{4-6}) that $n$ and $n_0$ are the quark numbers 
of $i=1$ and $i=2,3$ in $\ket{m_1}$, respectively. 
The eigenvalue equations for ${\hat n}$ and ${\hat n}_0$ give these numbers. 
Since ${\hat n}-{\hat n}_0={\hat a}^*{\hat a}+\sum_i{\hat b}_i^*{\hat b}_i$ and, 
then, 
$({\hat n}-{\hat n}_0)$ is positive definite, we have 
\beq\label{4-7-0}
n\geq n_0 \ .
\eeq
The operators ${\hat n}_0$ and ${\hat n}$ commute with the $su(4)$-generators 
and we have 
\bsub\label{4-7}
\beq
& &{\hat n}_0\ket{\lambda\rho\sigma_0\sigma}=n_0\ket{\lambda\rho\sigma_0\sigma} \ , 
\label{4-7a}\\
& &{\hat n}\ket{\lambda\rho\sigma_0\sigma}=n\ket{\lambda\rho\sigma_0\sigma} \ .
\label{4-7b}
\eeq
\esub
The relation (\ref{4-7}) suggests us that even if the eigenstates of ${\hat H}_m$ are not 
expressed explicitly in terms of the minimum weight states such as 
the form (\ref{4-4}), we can determine the quark 
numbers $n_0$ and $n$ charactering the minimum weight states.

As a possible expression of the eigenstates for ${\hat H}_m$, we presented the 
following form in (B):
\beq\label{4-8}
\dket{lsrw}=({\hat S}^3)^{2l}({\hat q}^1)^{2s}({\hat B}^*)^{2r}({\hat b}^*)^{2w}\ket{0} \ .
\eeq
Here, ${\hat S}^3$, ${\hat q}^1$ and ${\hat B}^*$ are defined in the relations (\ref{3-1}), 
(\ref{3-11}) and (\ref{3-12}), respectively. 
In (B), we proved that the expressions (\ref{4-4}) and (\ref{4-8}) are equivalent to each other 
under the correspondence 
\beq\label{4-9}
l=\lambda\ , \quad s=\sigma-\sigma_0-\rho \ , \quad
r=\rho \ , \quad w=\sigma \ .
\eeq
Since the operators ${\hat q}^i$, ${\hat S}^j$ and ${\hat B}^*$ commute 
with one another, the ordering of these operators in $\dket{lsrw}$ is 
arbitrary. 
As is shown in the relation (\ref{3-15}), ${\hat q}^1$, ${\hat S}^3$ and 
${\hat B}^*$ in the state $\dket{lsrw}$ carry one, two and three quarks. 
This indicates that the state (\ref{4-8}) is expressed in terms of 
the product of single-quarks, quark-pairs and quark-triplets. 
The above-mentioned structure of the state (\ref{4-8}) is quite 
interesting, and in this sense, the expression (\ref{4-8}) seems to be 
superior to the expression (\ref{4-4}). 
In (B), we showed that, in relation to the $su(1,1)$-algebra, 
the state (\ref{4-8}) is classified into two groups. 
First group obeys the condition $0 \leq r \leq (w-(s+l))/2$ and the second 
obeys $(w-(s+l))/2 \leq r \leq w-(s+l)$. 
They are shown in the relations (B$\cdot$5$\cdot$26) and (B$\cdot$5$\cdot$39), respectively. 
We are interested to investigate how the single-quarks, the quark-pairs and the quark-triplets coexist 
with one another in many-quark system. 
Therefore, irrelevantly to the magnitude of $s$ and $l$, it may be 
important to investigate the case $r=0$, in which there does not exist 
a quark-triplet. 
From the above reason, in this paper, we will investigate the case of the first group.

With the use of the relations (\ref{3-16}) $\sim$ (\ref{3-18}), we can prove the following relations: 
\bsub\label{4-10}
\beq
& &{\hat S}_2^1\dket{lsrw}={\hat S}_3^1\dket{lsrw}={\hat S}_3^2\dket{lsrw}=0 \ , 
\label{4-10a}\\
& &{\hat S}_1^1\dket{lsrw}=2(l+2r-w)\dket{lsrw} \ , 
\nonumber\\
& &{\hat S}_2^2\dket{lsrw}=2(l+s+2r-w)\dket{lsrw} \ , 
\nonumber\\
& &{\hat S}_3^3\dket{lsrw}=2(2l+s+2r-w)\dket{lsrw} \ . 
\label{4-10b}
\eeq
The state (\ref{4-8}) is the eigenstate of ${\hat n}_0$, ${\hat n}$, ${\hat N}$ and ${\hat {\mib Q}}^2$: 
\beq
& &{\hat n}_0\dket{lsrw}=n_0\dket{lsrw} \ , \ \qquad n_0=\Omega-w \ , 
\label{4-10c}\\
& &{\hat n}\dket{lsrw}=n\dket{lsrw} \ , \ \quad\qquad n=\Omega-w+2(s+r) \ , 
\label{4-10d}\\
& &{\hat N}\dket{lsrw}=N\dket{lsrw} \ , \ \ \qquad N=3(\Omega-w)+2(s+2l+3r) \ , 
\label{4-10e}\\
& &{\hat {\mib Q}}^2\dket{lsrw}=Q^2\dket{lsrw} \ , \qquad Q^2=
\frac{8}{3}(s^2+sl+l^2)+4(s+l) \ . 
\label{4-10f}
\eeq
\esub
With the use of the relation (\ref{4-9}), the eigenvalues of ${\hat n}_0$ and ${\hat n}$ become 
the forms shown in the relations (\ref{4-5a}) and (\ref{4-5b}), respectively. 
It may be interesting to see that the state (\ref{4-8}) is not expressed with 
the explicit use of $\ket{m_1}$, but, with the aid of ${\hat n}_0$ and ${\hat n}$, 
we can determine the eigenvalues of ${\hat n}_0$ and ${\hat n}$ 
characterizing $\ket{m_1}$ in the same results as that given in the state (\ref{4-4}). 
The relation (\ref{4-10}) tells us that $\dket{lsrw}$ is a minimum weight state of the 
$su(3)$-algebra based on the case 
(A$\cdot$4$\cdot$5).  
Further, $\dket{lsrw}$ is also a minimum weight state of the 
$su(1,1)$-algebra:
\beq\label{4-11}
{\hat T}_-\dket{lsrw}=0 \ , \qquad
{\hat T}_0\dket{lsrw}=(w+2)\dket{lsrw} \ . 
\eeq
With the use of the relations (\ref{4-10}) and (\ref{4-11}), we can construct the 
complete orthogonal set for the present system consisting of the 
eight kinds of bosons, which is shown in the relation (B$\cdot$4$\cdot$8).

The operation of ${\hat S}_1^2$, ${\hat S}_1^3$ and ${\hat S}_2^3$ on $\dket{lsrw}$ leads 
us to 
\beq\label{4-12}
& &{\hat S}_1^2\dket{lsrw}=-2s{\hat q}^2\dket{ls-r/2w} \ , \nonumber\\
& &{\hat S}_1^3\dket{lsrw}=2l{\hat S}^1\dket{l-s/2rw} -2s{\hat q}^3\dket{ls-r/2w}\ , 
\nonumber\\
& &{\hat S}_2^3\dket{lsrw}=2l{\hat S}^2\dket{l-s/2rw} \ .
\eeq
Combining the relations (\ref{4-10}) and (\ref{4-12}) with the condition (\ref{2-9}), 
we can learn that the state $\dket{lsrw}$ becomes color-singlet only in the case 
\beq\label{4-14}
l=s=0 \ , \quad {\rm i.e.,}\quad \dket{l=0\ s=0\ rw}=({\hat B}^*)^{2r}({\hat b}^*)^{2w}\ket{0} \ .
\eeq
The above indicates that only the states consisting of the quark-triplets are color-singlet. 
However, even if the condition (\ref{2-11}) is applied to $\dket{lsrw}$, the above 
conclusion does not change. 
Certainly, for any values of $l$ and $s$, we have 
\beq\label{4-15}
\dbra{lsrw}{\hat S}_i^j\dket{lsrw}=0\quad {\rm for}\quad i\neq j \ . 
\eeq 
However, for the expectation values of ${\hat S}_i^i$ for the normalized 
$\dket{lsrw}$, we have 
\beq\label{4-16}
& &\dbra{lsrw}{\hat S}_1^1\dket{lsrw}=2(l+2r-w) \ , \nonumber\\
& &\dbra{lsrw}{\hat S}_2^2\dket{lsrw}=2(l+s+2r-w) \ , \nonumber\\
& &\dbra{lsrw}{\hat S}_3^3\dket{lsrw}=2(2l+s+2r-w) \ .
\eeq
The relation (\ref{4-16}) tells us that except the case $l=s=0$, 
the expectation values are not equal. 
The above conclusion seems to suggest us that the color-singlet states 
cannot be described in terms of the states constructed on a chosen single minimum weight state, 
for example, such as $\ket{m_1}$. 
Of course, the case (\ref{4-14}) is an exception, i.e., the case $l=s=0$. 
If the above conclusion is reasonable, our next task is to investigate other types 
of the minimum weight states.

\section{Construction of the ``color-singlet" states based on the minimum weight states 
under the permutation for the color quantum numbers}

In this section, we will investigate the state $\ket{cs}_L$, the counterpart 
of $\rket{cs}_L$. 
As was shown in the relations (\ref{2-4}) and (\ref{2-5}), there exist 
six cases for fixing the $su(3)$-generators in the $su(4)$-algebra. 
The minimum weight state which we have adopted until the present 
is the state $\ket{m_1}$ given in the form (\ref{4-3}). 
It comes from one of the six cases, i.e., the case (\ref{2-4a}). 
In order to approach our problem for obtaining the ``color-singlet" states, 
it may be indispensable to examine other five cases.

The five minimum weight states are derived formally by the permutations for 
$i=1,\ 2$ and 3 from $\ket{m_1}$, and then, for a moment, we denote $\ket{m_1}$ 
as $\ket{m_{123}}$. 
Under this notation, the six minimum weight states can be expressed as 
$\ket{m_{i_1i_2i_3}}$. 
Here, $(i_1i_2i_3)$ is obtained from $(123)$ by the permutation 
$(1\rightarrow i_1, 2\rightarrow i_2, 3\rightarrow i_3)$. 
The case $(i_1=1, i_2=2, i_3=3)$ is identical 
permutation. 
Under the above notation, the minimum weight states coming from 
the cases (\ref{2-4a}) $\sim$ (\ref{2-4c}) are expressed as 
\bsub\label{5-1}
\beq
& &\ket{m_{123}}=({\hat b}_1^*)^{2(\sigma-\sigma_0)}
({\hat b}^*)^{2\sigma}\ket{0}\ (=\ket{m_1}) \ , 
\label{5-1a}\\
& &\ket{m_{231}}=({\hat b}_2^*)^{2(\sigma-\sigma_0)}
({\hat b}^*)^{2\sigma}\ket{0}\ (=\ket{m_2}) \ , 
\label{5-1b}\\
& &\ket{m_{312}}=({\hat b}_3^*)^{2(\sigma-\sigma_0)}
({\hat b}^*)^{2\sigma}\ket{0}\ (=\ket{m_3}) \ . 
\label{5-1c}
\eeq
\esub
The cases (\ref{2-5a}) $\sim$ (\ref{2-5c}) give us 
\beq\label{5-2}
\ket{m_{132}}=\ket{m_1} \ , \qquad
\ket{m_{213}}=\ket{m_2} \ , \qquad
\ket{m_{321}}=\ket{m_3} \ .
\eeq
The eigenstates of ${\wtilde H}_m$ are obtained by operating the 
$su(4)$-generators appropriately on the minimum weight states. 
Therefore, it may be enough to consider the states $\ket{m_1}$, 
$\ket{m_2}$ and $\ket{m_3}$. 
Of course,  $\ket{m_1}$, 
$\ket{m_2}$ and $\ket{m_3}$ satisfy  
\bsub\label{5-3}
\beq
& &{\hat S}_1\ket{m_1}={\hat S}_2\ket{m_1}={\hat S}_3\ket{m_1}=0 \ , 
\nonumber\\
& &{\hat S}_2^1\ket{m_1}={\hat S}_3^1\ket{m_1}={\hat S}_3^2\ket{m_1}=0 \ , 
\nonumber\\
& &{\hat S}_1^1\ket{m_1}=-2\sigma\ket{m_1} \ , \qquad
{\hat S}_2^2\ket{m_1}={\hat S}_3^3\ket{m_1}=-2\sigma_0\ket{m_1} \ , 
\label{5-3a}\\
& &\ \nonumber\\
& &{\hat S}_1\ket{m_2}={\hat S}_2\ket{m_2}={\hat S}_3\ket{m_2}=0 \ , 
\nonumber\\
& &{\hat S}_3^2\ket{m_2}={\hat S}_1^2\ket{m_2}={\hat S}_1^3\ket{m_2}=0 \ , 
\nonumber\\
& &{\hat S}_2^2\ket{m_2}=-2\sigma\ket{m_2} \ , \qquad
{\hat S}_3^3\ket{m_2}={\hat S}_1^1\ket{m_2}=-2\sigma_0\ket{m_2} \ , 
\label{5-3b}\\
& &\ \nonumber\\
& &{\hat S}_1\ket{m_3}={\hat S}_2\ket{m_3}={\hat S}_3\ket{m_3}=0 \ , 
\nonumber\\
& &{\hat S}_1^3\ket{m_3}={\hat S}_2^3\ket{m_3}={\hat S}_2^1\ket{m_3}=0 \ , 
\nonumber\\
& &{\hat S}_3^3\ket{m_3}=-2\sigma\ket{m_3} \ , \qquad
{\hat S}_1^1\ket{m_3}={\hat S}_2^2\ket{m_3}=-2\sigma_0\ket{m_3} \ . 
\label{5-3c}
\eeq
\esub
We will construct the orthogonal sets based on the minimum weight states 
(\ref{5-1a}) $\sim$ (\ref{5-1c}).

First, we present the minimum weight states for the $su(3)$-algebra which are 
orthogonal to $\dket{lsrw}$ shown in the form (\ref{4-8}). 
The state $\dket{lsrw}$ is constructed on the state $\ket{m_1}$\ ($=
\ket{m_{123}}$). 
Hereafter, we express the state $\dket{lsrw}$ in the following form: 
\bsub
\beq\label{5-4a}
\dket{123;slrw}&=&N_{slrw}({\hat q}^1)^{2s}({\hat S}^3)^{2l}
({\hat B}^*)^{2r}({\hat b}^*)^{2w}\ket{0} \nonumber\\
(&=&\dket{1;slrw}) \ . 
\eeq
With the aim of using the cyclic permutation $(1\rightarrow 2, 2\rightarrow 3, 
3\rightarrow 1)$, we changed the notation $\dket{lsrw}$ to 
$\dket{1;slrw}$, where $s$ and $l$ are transposed and $N_{slrw}$ denotes the 
normalization constant. 
By the cyclic permutation, we obtain other two states constructed on 
$\ket{m_2}\ (=\ket{m_{231}})$ and $\ket{m_3}\ (=\ket{m_{312}})$ in the 
form 
\beq
\dket{231;slrw}&=&N_{slrw}({\hat q}^2)^{2s}({\hat S}^1)^{2l}
({\hat B}^*)^{2r}({\hat b}^*)^{2w}\ket{0} \nonumber\\
(&=&\dket{2;slrw}) \ , 
\label{5-4b}\\
\dket{312;slrw}&=&N_{slrw}({\hat q}^3)^{2s}({\hat S}^2)^{2l}
({\hat B}^*)^{2r}({\hat b}^*)^{2w}\ket{0} \nonumber\\
(&=&\dket{3;slrw}) \ . 
\label{5-4c}
\eeq
\esub
In the case $s=l=0$, the above three states become identical: 
$\dket{i;s=0\ l=0\ rw}=N_{s=0l=0rw}({\hat B}^*)^{2r}({\hat b}^*)^{2w}\ket{0}$. 
It is color-singlet state $\ket{cs}$. 
Therefore, hereafter, we consider the cases except the case $s=l=0$. 
The above three states are the minimum weight states of the 
$su(3)$-algebra: 
\bsub\label{5-5}
\beq
& &{\hat S}_2^1\dket{1;slrw}={\hat S}_3^1\dket{1;slrw}={\hat S}_3^2\dket{1;slrw}=0\ , 
\label{5-5a}\\
& &{\hat S}_3^2\dket{2;slrw}={\hat S}_1^2\dket{2;slrw}={\hat S}_1^3\dket{2;slrw}=0\ , 
\label{5-5b}\\
& &{\hat S}_1^3\dket{3;slrw}={\hat S}_2^3\dket{3;slrw}={\hat S}_2^1\dket{3;slrw}=0\ , 
\label{5-5c}
\eeq
\esub
\vspace{-0.8cm}
\beq
& &{\hat S}_1^1\dket{1;slrw}=2(l+2r-w)\dket{1;slrw}\ , \qquad\quad\nonumber\\
& &{\hat S}_1^1\dket{2;slrw}=2(s+2l+2r-w)\dket{2;slrw}\ ,\nonumber\\
& &{\hat S}_1^1\dket{3;slrw}=2(s+l+2r-w)\dket{3;slrw}\ ,  
\label{5-6}\\
& &\ \nonumber\\
& &{\hat S}_2^2\dket{1;slrw}=2(s+l+2r-w)\dket{1;slrw}\ , \nonumber\\
& &{\hat S}_2^2\dket{2;slrw}=2(l+2r-w)\dket{2;slrw}\ ,\nonumber\\
& &{\hat S}_2^2\dket{3;slrw}=2(s+2l+2r-w)\dket{3;slrw}\ ,  
\label{5-7}\\
& &\ \nonumber\\
& &{\hat S}_3^3\dket{1;slrw}=2(s+2l+2r-w)\dket{1;slrw}\ , \nonumber\\
& &{\hat S}_3^3\dket{2;slrw}=2(s+l+2r-w)\dket{2;slrw}\ ,\nonumber\\
& &{\hat S}_3^3\dket{3;slrw}=2(l+2r-w)\dket{3;slrw}\ .  
\label{5-8}
\eeq
The relation (\ref{5-6}) $\sim$ (\ref{5-8}) tell us that the states are 
orthogonal to one another. 
With the use of the relations (\ref{5-5a}) $\sim$ (\ref{5-5c}), we can 
prove that all the matrix elements except the following ones vanish: 
\beq\label{5-9}
& &\dbra{2;s'l'r'w'}{\hat S}_1^i\dket{1;slrw}\quad {\rm for} \quad i=2,\ 3\ , 
\nonumber\\
& &\dbra{3;s'l'r'w'}{\hat S}_2^i\dket{2;slrw}\quad {\rm for} \quad i=3,\ 1\ , 
\nonumber\\
& &\dbra{1;s'l'r'w'}{\hat S}_3^i\dket{3;slrw}\quad {\rm for} \quad i=1,\ 2\ . 
\eeq
However, these matrix elements also vanish. 
For example, 
${\hat S}_1^i\dket{1;slrw}$ is not constructed by operating 
${\hat S}_1^i$ on the state $\dket{2;slrw}$ and this indicates that it does not 
belong to the irreducible representation based on the minimum weight state 
$\dket{2;slrw}$. 
From the above argument, the matrix element 
$\dbra{2;s'l'r'w'}{\hat S}_1^i\dket{1;slrw}$ vanishes. 
Other cases are also in the same situation as the above one. 
The above minimum weight states are the eigenstates of ${\hat n}_0$ and ${\hat n}$ with 
the eigenvalues $n_0=\Omega-w=\Omega-\sigma$ and $n=\Omega-w+2(s+r)=\Omega+\sigma-2\sigma_0$, 
respectively. 
These values are shown in the relations (\ref{4-10c}) and (\ref{4-10d}). 
The reason is very simple: 
The operators ${\hat n}_0$ and ${\hat n}$ are color-symmetric.

Under the above consideration, we will construct the ``color-singlet" states in the 
present model. 
In the case except the case $s=l=0$, it is impossible to construct the 
states obeying the condition (\ref{2-9}). 
Then, we take up the condition (\ref{2-11}). 
First, we set up a state which is a linear combination for the states 
$\dket{p;slrw}$ for $p=1,\ 2$ and 3: 
\beq\label{5-10}
\dket{cs;slrw}=\sum_p C_p(slrw)\dket{p;slrw} \ . 
\eeq
Here, $C_p(slrw)$ denotes the coefficient of the linear combination 
with $\sum_p|C_p(slrw)|^2=1$. 
We note that for any values of $C_p(slrw)$, the state (\ref{5-10}) satisfies 
\bsub\label{5-11}
\beq
\dbra{cs;slrw}{\hat S}_i^j\dket{cs;slrw}=0\quad {\rm for}\quad i\neq j \ . 
\label{5-11a}
\eeq
The relation (\ref{5-11a}) is nothing but the condition (\ref{2-11a}). 
Concerning the condition (\ref{2-11b}), we are able to have 
\beq\label{5-11b}
& &\dbra{cs;slrw}{\hat S}_1^1\dket{cs;slrw}=\dbra{cs;slrw}{\hat S}_2^2\dket{cs;slrw}\nonumber\\
&=&\dbra{cs;slrw}{\hat S}_3^3\dket{cs;slrw}=\frac{4}{3}(s+2l)+2(2r-w) \ .
\eeq
\esub
The relation (\ref{5-11b}) is realized in the case 
\beq\label{5-12}
|C_p(slrw)|^2=\frac{1}{3} \ . 
\eeq
Any case except the relation (\ref{5-12}) does not lead to the condition (\ref{2-11b}). 
In the present framework, it is impossible to fix the phase factor of 
$C_p(slrw)$, but, it may be enough to adopt the following form: 
\beq\label{5-13}
\ket{cs}_L&=&
\dket{cs;slrw}=\frac{1}{\sqrt{3}}\sum_p \dket{p;slrw} \nonumber\\
&=&\frac{1}{\sqrt{3}}(\dket{123;slrw}+\dket{231;slrw}+\dket{312;slrw}) \ .
\eeq
It may be permitted to say that the state $\ket{cs}_L=\dket{cs;slrw}$ is color-symmetric 
state with respect to the cyclic permutation for $p=1,2$ and 3. 
Since ${\hat n}_0$, ${\hat n}$ and ${\hat N}$ are color-symmetric, 
the results obtained in the state $\dket{lsrw}$ are also found in the state 
$\dket{cs;slrw}$. 
The results are shown in the relations (\ref{4-10c}) $\sim$ (\ref{4-10f}). 
The energy eigenvalue is also in the same situation as the above. 
The details will be discussed in the next section. 
The above argument supports that any information related to the 
eigenvalue of color-symmetric operator given in (A) $\sim$ (C) is also valid in the 
``color-singlet" state $\ket{cs}_L$.

At the end of \S 2, we mentioned that it may be desirable to search the state 
$\rket{cs}_L$, in which $\rbra{cs}{\wtilde {\mib Q}}^2\rket{cs}_L$ is 
as small as possible, i.e., $\rket{cs}_M$. 
We complement this statement. 
There does not exist the absolute condition for 
the state $\rket{cs}_M$ which we intend to search. 
In the next section, we will give a possible idea for 
this condition.

\section{Construction of the ``color-singlet" states minimizing the eigenvalue of the 
$su(3)$-Casimir operator}

In this section, we will investigate the state $\ket{cs}_M$, the counterpart of 
$\rket{cs}_M$. 
In the relation (\ref{5-13}), we presented the ``color-singlet" 
states $\ket{cs}_L$ which obey the condition (\ref{2-11}). 
After discussing some characteristic features produced by these states, 
we will formulate the states $\ket{cs}_M$ in our idea. 
As was already mentioned in \S 4, the expressions (\ref{4-4}) and (\ref{4-8}) 
are equivalent to each other through the relation (\ref{4-9}) for the 
quantum numbers specifying the states. 
The relation (\ref{4-9}) is also valid for the case of the 
states $\ket{cs}_L$. 
In (A), after lengthy argument, we showed the overall ranges in which the 
quantum numbers can change. 
In this paper, we investigate the range based on the first group mentioned in \S 4.  
 
First, we notice the following relation: 
\bsub\label{6-1}
\beq
& &2s \geq 0 \ , \qquad 2l \geq 0 \ , \qquad 2r \geq 0 \ , \qquad 2w \geq 0 \ , 
\label{6-1a}\\
& &2s+2l+4r \leq 2w \ . 
\label{6-1b}
\eeq
\esub
The relation (\ref{6-1a}) comes from the condition that the exponents 
$2s$, $2l$, $2r$ and $2w$ in the relation (\ref{5-4a}) should be positive or zero. 
Strictly speaking, the quantum numbers $2s$, $2l$, $2r$ and $2w$ are positive integers. 
However, for simplicity, we will treat them as continuously varying positive parameters. 
The relation (\ref{6-1b}) comes from the first group for $\dket{cs;slrw}$. 
It should be noted that $2s$, $2l$ and $2r$ denote the numbers of the 
single-quarks, the quark-pairs and the quark-triplets, respectively. 
Further, we note the eigenvalues of the eigenvalue equations 
(\ref{4-10c})$\sim$(\ref{4-10e}), which lead to 
\bsub\label{6-2}
\beq
& &w=\Omega^0 \ , 
\label{6-2a}\\
& &2s+2r=n^0\ , \qquad
2l+2r=\frac{1}{2}(N^0-n^0) \ . 
\label{6-2b}
\eeq
The relations (\ref{6-2a}) and (\ref{6-2b}) give us
\beq\label{6-2c}
(2s+2l+4r)-2w=\frac{1}{2}(N^0+n^0)-2\Omega^0\ . 
\eeq
\esub
Here, $\Omega^0$, $n^0$ and $N^0$ are defined as 
\beq\label{6-3}
\Omega^0=\Omega-n_0 \ , \qquad
n^0=n-n_0 \ , \qquad
N^0=N-n_0 \ .
\eeq
The quantities $\Omega^0$, $n^0$ and $N^0$ are also regarded as continuously 
varying positive parameters. 
Combining the relation (\ref{6-2}) with the inequality 
(\ref{6-1}), we have the inequality. 
\begin{figure}[t]
\begin{center}
\includegraphics[height=5.5cm]{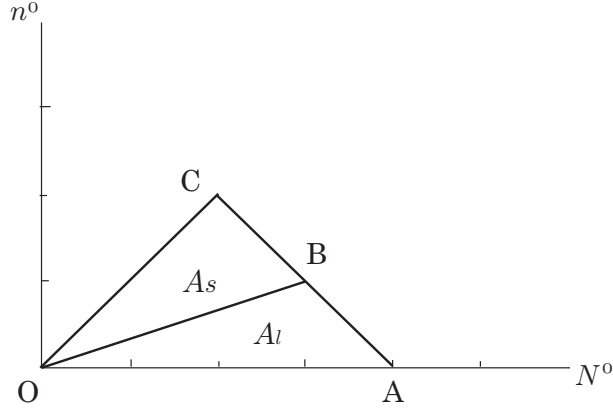}
\caption{
Two areas in which $n^0$ and $N^0$ satisfy for a given value $\Omega^0$ 
are depicted.
The line OA represents $n^0=0$, the line OB represents $n^0=N^0/3$, 
the line OC represents $n^0=N^0$ and the line AC represents 
$n^0=4\Omega^0-N^0$. 
}
\label{fig:6-1}
\end{center}
\end{figure}
\beq\label{6-4}
0 \leq n^0 \ , \qquad 
n^0 \leq N^0 \ , \qquad
n^0 \leq 4\Omega^0-N^0 \ . 
\eeq
The relation (\ref{6-2b}) gives us 
\beq\label{6-5}
2s-2l=\frac{3}{2}\left(n^0-\frac{N^0}{3}\right) \ .
\eeq
From the relation (\ref{6-5}), we have the following relation: 
\bsub\label{6-6}
\beq
& &{\rm if}\qquad 2s \leq 2l \ , \qquad n^0\leq \frac{N^0}{3} \ , 
\label{6-6a}\\
& &{\rm if}\qquad 2s \geq 2l \ , \qquad n^0 \geq \frac{N^0}{3} \ . 
\label{6-6b}
\eeq
\esub
Figure 1 shows two areas $A_l$ and $A_s$, where the 
inequalities (\ref{6-4}) and (\ref{6-6}) are satisfied. 
The line OA gives $2s=2r=0$ and $2l=N^0/2$, in which the system 
consists of only $N^0/2$ quark-pairs. 
On the line OC, we have $2l=2r=0$ and $2s=N^0$. 
In this case, the system consists of only $N^0$ single-quarks. 
On the line OB, we have $2s=2l=N^0/3-2r$. 
In this case, the numbers of the single-quarks and the quark-pairs are 
equal to each other. 
If $2s=2l=0$, $2r=N^0/3$ and this case corresponds to the quark-triplets 
and, of course, if $2r=0$, we have 
$2s=2l=N^0/3$. 
In the area $A_l$, the number of the quark-pairs is larger than that of the 
single-quarks and in the area $A_s$, the vice versa. 
The relation (\ref{6-2b}) gives us $2s=n^0-2r \geq 0$ and 
$2l=(N^0-n^0)/2-2r\geq 0$ and, 
then, we can show that in the areas $A_l$ and $A_s$, $2r$ can 
changes its value in the following ranges, respectively: 
\bsub\label{6-7}
\beq
& &A_l\ : \ \ 0 \leq 2r \leq n^0 \ , 
\label{6-7a}\\
& &A_s \ : \ \ 0\leq 2r \leq \frac{1}{2}(N^0-n^0) \ . 
\label{6-7b}
\eeq
\esub
Later, the relation (\ref{6-7}) will play a central role.

In Fig.1, we see that $N^0$ changes its value in the range 
\beq\label{6-8}
0 \leq N^0 \leq 4\Omega^0 \ .
\eeq
However, the range of $N$ in the present model is as follows: 
\beq\label{6-9}
3n^0 \leq N \leq 6\Omega-3n^0 \ , \quad {\rm i.e.,}\quad
0 \leq N^0 \leq 6\Omega^0 \ .
\eeq
Therefore, we must interpret the discrepancy between $4\Omega^0$ and $6\Omega^0$. 
As was discussed in \S 2.3 in (A), we can make re-formation of the present model 
from the side of $N=6\Omega$, which we called the hole picture in (A). 
If we follow this re-formation, it is enough to replace $N^0$ with 
$(6\Omega^0-N^0)$ in the relation obtained as function of $N^0$ in the 
present form: 
\beq\label{6-10}
2\Omega^0 \leq N^0 \leq 6\Omega^0 \ , \quad
0 \leq n^0 \ , \quad
n^0\leq 6\Omega^0-N^0 \ , \quad
n^0\leq -2\Omega^0+N^0 \ . 
\eeq
The line $n^0=N^0/3$ changes to $n^0=2\Omega^0-N^0/3$. 
Figure 2 shows the areas given in the relation (\ref{6-10}), 
together with the areas given in the relation (\ref{6-4}). 
We can see that both forms cover the whole ranges.

\begin{figure}[t]
\begin{center}
\includegraphics[height=5.5cm]{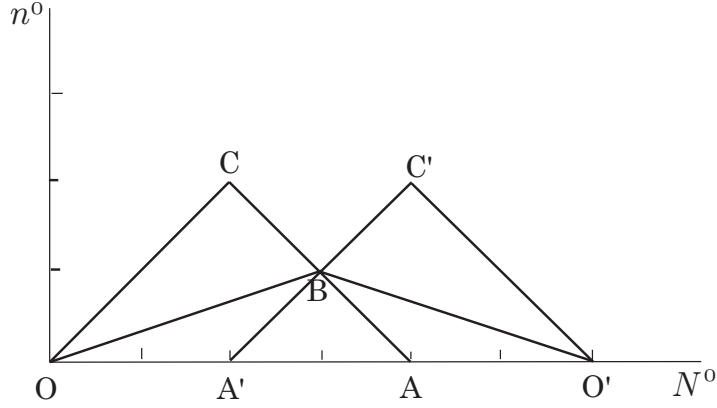}
\caption{
New triangle A'O'C' is obtained from old triangle AOC by replacing $N^0$ with 
$(6\Omega-N^0)$. 
}
\label{fig:6-3}
\end{center}
\end{figure}

The eigenvalue of the $su(3)$-Casimir operator, $Q^2$, is shown 
in the relation (\ref{4-10f}). 
It is expressed in terms of $2s$ and $2l$ and using the relation 
(\ref{6-2b}), it can be expressed in terms of 
$N^0$, $n^0$ and $2r$. 
Hereafter, we denote $Q^2$ as $F(N^0,n^0;2r)$: 
\beq\label{6-11-0}
F(N^0,n^0;2r)=2\left[2r-\left(\frac{1}{4}(N^0+n^0)+1\right)\right]^2
+\frac{3}{8}\left[\left(n^0-\frac{N^0}{3}\right)^2-\frac{16}{3}\right] \ . \ \ 
\eeq
Since $Q^2$ is positive-definite, for given values of $N^0$ and $n^0$, 
$2r$ is meaningful in the regions satisfying the condition 
$F(N^0,n^0;2r)\geq 0$. 
For given values of $N^0$ and $n^0$, they appear in the regions 
\bsub\label{6-11}
\beq
& &(1)\ \ {\rm If} \ \ n^0 \geq \frac{N^0}{3}+\frac{4}{\sqrt{3}}\ \ {\rm or}\ \ 
n^0 \leq \frac{N^0}{3}-\frac{4}{\sqrt{3}} \ , \nonumber\\
& &\qquad\qquad\qquad 
-\infty < 2r < +\infty \ , 
\label{6-11a}\\
& &(2)\ \ {\rm If} \ \ \frac{N^0}{3}-\frac{4}{\sqrt{3}} \leq n^0 \leq \frac{N^0}{3}+\frac{4}{\sqrt{3}}\ ,  
\nonumber\\
& &\qquad\qquad\qquad 
2r \leq \frac{1}{4}(N^0+n^0)+1-\sqrt{1-\frac{3}{16}\left(n^0-\frac{N^0}{3}\right)^2} \ , 
\label{6-11b}\\
& &\qquad\qquad\qquad
2r \geq \frac{1}{4}(N^0+n^0)+1+\sqrt{1-\frac{3}{16}\left(n^0-\frac{N^0}{3}\right)^2} \ . 
\label{6-11c}
\eeq
\esub
In the case (1), we can prove 
\bsub\label{6-12}
\beq
& &n^0 \leq \frac{1}{4}(N^0+n^0)+1 \qquad\qquad\quad {\rm for}\quad 
n^0\leq \frac{N^0}{3}-\frac{4}{\sqrt{3}} \ , 
\label{6-12a}\\
& &\frac{1}{2}(N^0-n^0) \leq \frac{1}{4}(N^0+n^0)+1 \quad {\rm for}\quad 
n^0 \geq \frac{N^0}{3}+\frac{4}{\sqrt{3}} \ .
\label{6-12b}
\eeq
\esub
In the case (2), we can prove 
\bsub\label{6-13}
\beq
& &n^0 \leq \frac{1}{4}(N^0+n^0)+1-\sqrt{1-\frac{3}{16}\left(n^0-\frac{N^0}{3}\right)^2}
\nonumber\\
& &\qquad\qquad\qquad\qquad\qquad\qquad\qquad\qquad
{\rm for}\quad 
\frac{N^0}{3}-\frac{4}{\sqrt{3}} \leq n^0 \leq \frac{N^0}{3} \ , 
\label{6-13a}\\
& &\frac{1}{2}(N^0-n^0) \leq \frac{1}{4}(N^0+n^0)+1-\sqrt{1-\frac{3}{16}\left(n^0-\frac{N^0}{3}\right)^2} 
\nonumber\\
& &\qquad\qquad\qquad\qquad\qquad\qquad\qquad\qquad
{\rm for}\quad 
\frac{N^0}{3} \leq n^0 \leq \frac{N^0}{3}+\frac{4}{\sqrt{3}} \ .
\label{6-13b}
\eeq
\esub
The function $F(N^0,n^0;2r)$ is monotone-decreasing in the range 
$0\leq 2r \leq (N^0+n^0)/4+1$. 
Therefore, the maximum values of $2r$, $2r_m$, are given as follows: 
\bsub\label{6-14}
\beq
& &2r_m=n^0 \quad\qquad\qquad {\rm in} \ {\rm A}_l \ , 
\label{6-14a}\\
& &2r_m=\frac{1}{2}(N^0-n^0) \quad {\rm in} \ {\rm A}_s \ . 
\label{6-14b}
\eeq
\esub
The behavior of $F(N^0,n^0;2r)$ leads to the following results:
The minimum values of $F(N^0,n^0;2r)$ are given 
by $F(N^0,n^0;2r_m)$ for given values of $N^0$ and $n^0$ and 
their values are explicitly calculated in the forms 
\bsub\label{6-15}
\beq
& &F(N^0,n^0,2r_m=n^0)=\frac{1}{6}(N^0-3n^0)^2+(N^0-3n^0) \qquad\quad
{\rm for}\ {\rm A}_l \ , 
\label{6-15a}\\
& &F(N^0,n^0,2r_m=\frac{1}{2}(N^0-n^0))=\frac{1}{6}(N^0-3n^0)^2-(N^0-3n^0) \quad
{\rm for}\ {\rm A}_s \ . \quad
\label{6-15b}
\eeq
\esub
In the forthcoming paper (Part II), we use the notations for $F(N^0,n^0,2r_m)$ and the 
``color-singlet" states $\ket{cs}_M$ leading to $F(N^0,n^0;2r_m)$ in the form 
$F_M(N^0,n^0)$ and $\ket{cs;N^0,n^0}_M$, respectively. 
From the above argument, it may be natural to regard $\ket{cs;N^0,n^0}_M$ as the 
``color-singlet" states $\ket{cs}_M$ which we have intended to search in this paper, 
because the states $\ket{cs;N^0,n^0}_M$ minimize $\bra{cs}{\hat {\mib Q}}^2\ket{cs}_L$ 
in the areas $A_l$ and $A_s$. 
It satisfies our requirement mentioned in \S 2. 
Of course, we can give the explicit expression for $\ket{cs;N^0,n^0}_M$, 
but it may be not necessary for present argument. 

In this paper, we presented the color-symmetric form of the $su(4)$-algebraic model 
for many-quark system. 
With the aid of the Schwinger boson realization, this form automatically leads to 
the ``color-singlet" states in the boson space, which satisfy the relation 
equivalent to the relation (\ref{2-11}) in the original fermion space. 
The color-symmetric hermitian operators, ${\hat {\mib P}}^2$, 
${\hat {\mib Q}}^2$, ${\hat N}$ and ${\hat n}$ are essential for the form. 
In various effective theories of QCD, the condition such as the relation (\ref{2-11}) 
plays a central role. 
However, in this paper, we pointed out that it may be sufficient to insure the color-singlet 
property, only in the frame of the condition (\ref{2-11}). 
Further, we stressed that, in addition to the condition (\ref{2-11}), the 
expectation value of ${\hat {\mib Q}}^2$ should be as small as possible. 
In the forthcoming paper (Part II), we will give various analysis for 
this condition.

In order to complete the present paper (Part I), we will give the expression of the 
energy eigenvalues. 
With the use of the expression (\ref{2-8b}), we can calculate the energy eigenvalues of 
${\hat H}_m$ for the states $\dket{cs;slrw}$. 
We know that the eigenvalues of ${\hat {\mib Q}}^2$ and ${\hat \Sigma}$ are 
expressed in the forms 
$F(N^0,n^0;2r)$ and $(2\Omega^0-N^0)(3\Omega^0-N^0+6)/3$, respectively. 
The relations (A$\cdot$2$\cdot$2b) and (\ref{4-5}) give us the eigenvalue of 
${\hat {\mib P}}^2$ in the form 
$(3\Omega^{02}-2\Omega^0n^0+n^{02}+6\Omega^0)$. 
Then, the energy eigenvalue of ${\hat H}_m$, which we denote as 
$E^{(m)}(N^0,n^0;2r)$, is obtained as follows:
\beq\label{7-1}
E^{(m)}(N^0,n^0;2r)&=&
\frac{1}{2}(1+2\chi)F(N^0,n^0;2r) \nonumber\\
& &+\frac{1}{2}n^0(2\Omega^0-n^0)-\frac{1}{6}N^0(6\Omega^0+6-N^0) \ . 
\eeq
Of course, the expression (\ref{7-1}) is given for the case 
$0 \leq N^0 \leq 4\Omega^0$. 
For the case $2\Omega^0 \leq N^0 \leq 6\Omega^0$, we have the 
expression for the energy eigenvalue, which we denote as 
${\cal E}^{(m)}(N^0,n^0;2r)$, in the form 
\beq\label{7-2}
{\cal E}^{(m)}(N^0,n^0;2r)=
E^{(m)}(6\Omega^0-N^0, n^0;2r)+2(3\Omega^0-N^0) \ .
\eeq
The above expression is obtained by replacing $N^0$ with $(6\Omega^0-N^0)$ 
in $E^{(m)}(N^0,n^0;2r)$ and by adding the term 
$2(3\Omega^0-N^0)$, which comes from the difference of the ordering of 
${\hat S}^i{\hat S}_i$ and ${\hat S}_i{\hat S}^i$. 
The detail has been discussed in (A). 
Hereafter, we will not contact with the case $2\Omega^0 \leq N^0 \leq 6\Omega^0$ 
explicitly. 
Of course, in the case where we discuss some aspects of the present model, 
we will use the results for $2\Omega^0\leq N^0 \leq 6\Omega^0$, if necessary.

By substituting the relation (\ref{6-15}) into the expression (\ref{7-1}), 
we obtain the energy eigenvalue for the state $\ket{cs;N^0,n^0}_M$: 
\bsub\label{7-3}
\beq
& &
{\rm For\ the\ area}\ A_l\ , \nonumber\\
& &\qquad
E_l^{(m)}(N^0,n^0)=E^{(m)}(N^0,n^0;2r_m) \nonumber\\
& &\qquad\qquad\qquad\qquad
=\frac{1}{4}(1+6\chi)n^{02}-\frac{1}{2}\left[
(N^0+3-2\Omega^0)+2\chi(N^0+3)\right]n^0 \nonumber\\
& &\qquad\qquad\qquad\qquad
\ -\frac{1}{4}N^0(4\Omega^0-N^0+2)+\frac{\chi}{6}N^0(N^0+6)\ , 
\label{7-3a}\\
& &
{\rm For\ the\ area}\ A_s\ , \nonumber\\
& &\qquad
E_s^{(m)}(N^0,n^0)=E^{(m)}(N^0,n^0;2r_m) \nonumber\\
& &\qquad\qquad\qquad\qquad
=\frac{1}{4}(1+6\chi)n^{02}-\frac{1}{2}\left[
(N^0-3-2\Omega^0)+2\chi(N^0-3)\right]n^0 \nonumber\\
& &\qquad\qquad\qquad\qquad
\ -\frac{1}{4}N^0(4\Omega^0-N^0+6)+\frac{\chi}{6}N^0(N^0-6) \ . 
\label{7-3b}
\eeq
\esub
In the discussion in Part II, the expressions (\ref{7-1}) and (\ref{7-2}) 
will play a central role. 
Through the discussion, we may be able to understand how the 
condition required for the expectation value of ${\hat {\mib Q}}^2$ will work in 
the energies.

\section{Summary}

In this paper, we investigated the ``color-singlet" state in the modified Bonn quark model 
by means of the boson realization. 
In the exact energy eigenstates, a color-neutral quark-triplet state is, 
of course, a color-singlet state under the ordinary 
color-singletness. 
However, we loosened this condition in terms of expectation values. 
In addition to the above-mentioned condition, we imposed a condition 
that the ``color-singlet" state minimizes the eigenvalue 
of the $su(3)$-Casimir operator. 
Then, we could construct the state as the superposition of the 
eigenstates constructed on the minimum weight states of some 
$su(3)$-subalgebras included in the dynamical $su(4)$-symmetry which the 
original Bonn quark model has.
Further, we analyzed the regions which consist of the single-quarks, 
the quark-pairs and the quark-triplets. 
The implication of the results on the color superconducting phase or nuclear matter 
phase in dense quark or nuclear matter will be investigated as one of future problems.


\appendix

\section{The $su(3)$-Algebra}

Usually, the generators of the $su(3)$-algebra are given as the 
following eight operators:
\beq\label{a1}
{\wtilde I}_{\pm,0} \ , \quad {\wtilde M} \ , \quad
{\wtilde D}_{\pm,}^* \ , \quad
{\wtilde D}_{\pm} \  .
\eeq
They obey the commutation relations 
\bsub\label{a2}
\beq
& &[\ {\wtilde I}_+\ , \ {\wtilde I}_- \ ]=2{\wtilde I}_0 \ , \qquad
[\ {\wtilde I}_0\ , \ {\wtilde I}_{\pm} \ ]=\pm{\wtilde I}_{\pm} \ ,
\label{a2a}\\
& &[\ {\wtilde I}_{\pm,0}\ , \ {\wtilde M}_0 \ ]=0 \ ,
\label{a2b}\\
& &[\ {\wtilde I}_{\pm}\ , \ {\wtilde D}_{\pm}^* \ ]=0 \ , \qquad
[\ {\wtilde I}_{\pm}\ , \ {\wtilde D}_{\mp}^* \ ]={\wtilde D}_{\pm}^* \ , 
\qquad
[\ {\wtilde I}_{0}\ , \ {\wtilde D}_{\pm}^* \ ]=\pm\frac{1}{2}{\wtilde D}_{\pm}^* \ ,
\label{a2c}\\
& &[\ {\wtilde M}_{0}\ , \ {\wtilde D}_{\pm}^* \ ]=\frac{3}{2}{\wtilde D}_{\pm}^* \ ,
\label{a2d}\\
& &[\ {\wtilde D}_{+}^*\ , \ {\wtilde D}_{-}^* \ ]=0 \ , \qquad
[\ {\wtilde D}_{\pm}^*\ , \ {\wtilde D}_{\pm} \ ]={\wtilde M}_{0}\pm {\wtilde I}_0 \ , 
\qquad
[\ {\wtilde D}_{\pm}^*\ , \ {\wtilde D}_{\mp}^* \ ]={\wtilde I}_{\pm} \ .
\label{a2e}
\eeq
\esub
The relation (\ref{a2a}) tells that $({\wtilde I}_{\pm,0})$ forms the 
$su(2)$-algebra. 
In the relations (\ref{a2b}) and (\ref{a2c}), we learn that ${\wtilde M}_0$ 
is scalar and ${\wtilde D}_{\pm}^*$ and ${\wtilde D}_{\mp}$ are spinor 
for $({\wtilde I}_{\pm,0})$.

In the case (\ref{2-4a}), the eight $su(3)$-generators can be expressed in the form 
\bsub\label{a3}
\beq
& &{\wtilde I}_+={\wtilde S}_2^3 \ , \qquad
{\wtilde I}_-={\wtilde S}_3^2\ , \qquad
{\wtilde I}_0=\frac{1}{2}({\wtilde S}_2^2-{\wtilde S}_3^3) \ , 
\label{a3a}\\
& &{\wtilde M}_0={\wtilde S}_1^1-\frac{1}{2}({\wtilde S}_2^2+{\wtilde S}_3^3) \ , 
\label{a3b}\\
& &{\wtilde D}_+^*={\wtilde S}_1^3 \ , \qquad
{\wtilde D}_-^*={\wtilde S}_1^2 \ , \qquad
{\wtilde D}_+={\wtilde S}_3^1 \ , \qquad
{\wtilde D}_-={\wtilde S}_2^1 \ . 
\label{a3c}
\eeq
\esub
The other cases are obtained under the permutation $(1\rightarrow i_1, 
2\rightarrow i_2, 3\rightarrow i_3)$.


\begin{thebibliography}{99}
\bibitem{01}
M. G. Alford, A. Schmitt, K. Rajagopal and T. Sch\"afer, Rev. Mod. Phys. {\bf 80} 
(2008), 1455, and references cited therein.
\bibitem{02}
K. Bleuler, H Hofest\"adt, S. Merk and H. R. Petry, Z. Naturforsch. {\bf 38a} 
(1983) 705.\\
H. R. Petry, in {\it Lecture Notes in Physics}, Vol. 197, ed. 
K. Bleuler (Springer, Berlin, 1984), p.236.\\
H. R. Petry, H. Hofest\"adt, S. Merk, K.Bleuler, H. Bohr and K. S. Narain, 
Phys. Lett. B {\bf 159} (1985), 363. 
\bibitem{1}
Y. Tsue, C. Provid\^encia, J. da Provid\^encia and M. Yamamura, Prog. Theor. Phys. 
{\bf 121} (2009), 1237. 
\bibitem{2}
Y. Tsue, C. Provid\^encia, J. da Provid\^encia and M. Yamamura, 
Prog. Theor. Phys. {\bf 122} (2009), 693: Errata, ibid {\bf 122} (2009), 1065.
\bibitem{3}
Y. Tsue, C. Provid\^encia, J. da Provid\^encia and M. Yamamura, Prog. Theor. Phys. 
{\bf 122} (2009), 911. 
\bibitem{b}
M. Yamamura, A. Kuriyama and T. Kunihiro, 
Prog. Theor. Phys. {\bf 104} (2000), 385. 
\end{thebibliography}
\end{document}